\documentclass[12pt]{article}
\usepackage{putex}
\usepackage{graphicx}

\begin{document}

\preprint{PUPT-2257}

\institution{PU}{Joseph Henry Laboratories, Princeton University, Princeton, NJ 08544}

\title{Bulk viscosity of strongly coupled plasmas with holographic duals}

\authors{Steven S. Gubser,  Silviu S. Pufu, and F\'abio D. Rocha}

\abstract{We explain a method for computing the bulk viscosity of strongly coupled thermal plasmas dual to supergravity backgrounds supported by one scalar field.  Whereas earlier investigations required the computation of the leading dissipative term in the dispersion relation for sound waves, our method requires only the leading frequency dependence of an appropriate Green's function in the low-frequency limit.  With a scalar potential chosen to mimic the equation of state of QCD, we observe a slight violation of the lower bound on the ratio of the bulk and shear viscosities conjectured in \cite{Buchel:2007mf}.

}

\date{June 2008}

\maketitle

\tableofcontents

\section{Introduction and summary}

Recent interest has attached to the possibility of a dramatic rise in the bulk viscosity $\zeta$ near the cross-over temperature $T_c$ of QCD where confinement and chiral symmetry breaking set in \cite{Kharzeev:2007wb,Karsch:2007jc}.  Bulk viscosity is accessible via AdS/CFT if we break conformal symmetry, for example by adding one or more scalars with non-zero profiles in the bulk.  Bulk viscosity was studied in a holographic setting in \cite{Parnachev:2005hh,Benincasa:2005iv} and subsequent work includes \cite{Buchel:2005cv,Benincasa:2006ei,Mas:2007ng,Buchel:2007mf}.  Recently, we showed in collaboration with A.~Nellore \cite{Gubser:2008yx} that if a single scalar is coupled to gravity and a potential is chosen for it to mimic the QCD equation of state, then the bulk viscosity rises significantly near $T_c$, but not as dramatically as expected for pure Yang-Mills theory based on the works \cite{Kharzeev:2007wb,Meyer:2007dy}.  The numerical studies in \cite{Gubser:2008yx} led to the conjecture that the only way to get $\zeta/s$ to diverge in the context of a holographic dual in the two-derivative supergravity approximation is for the entropy density $s$ to have an extremum as a function of temperature.

\begin{figure}
   \centering
   \includegraphics[width=5in]{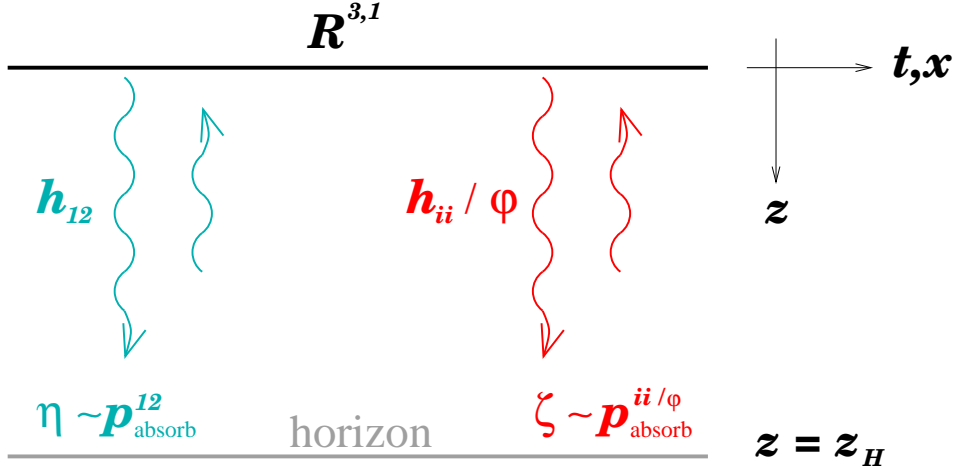}
   \caption{A cartoon of the relation between shear viscosity and $h_{12}$ graviton absorption, and between bulk viscosity and absorption of a mixture of the $h_{ii}$ graviton and the scalar $\phi$.} \label{ViscosityVsProability}
 \end{figure}

The purpose of this paper is to provide an exposition of the methods used for calculating the bulk viscosity in finite-temperature gravity duals involving a single scalar, as explored for example in \cite{Gubser:2008ny,Gursoy:2008bu}.  Earlier works \cite{Parnachev:2005hh,Benincasa:2005iv,Buchel:2005cv,Benincasa:2006ei,Mas:2007ng,Buchel:2007mf} have focused on calculating the leading dissipative term in the dispersion relation for sound waves.  We instead appeal directly to a Kubo formula.  Our computation comes down to understanding the bulk propagation of some linear superposition of gravitons and scalar perturbations.  The strategy, as illustrated in figure~\ref{ViscosityVsProability}, is to start with an infinite, static, thermal background, linearize the equations of motion around it, and solve these linearized equations in some appropriate approximation.
A large-wavelength, low-frequency approximation is the appropriate one because $\zeta$ describes an effect in hydrodynamics, which is to say the infrared approximation to the dynamics.  The absorption probability of low-energy quanta is proportional to $\zeta$.  This is entirely analogous to the well-known computation of shear viscosity \cite{Policastro:2001yc}, but more technically involved because of the mixing between the scalar and metric perturbations. We can decouple these perturbations through a judicious choice of gauge, namely the gauge where the value of the scalar is used as a radial variable. In this gauge, we need only solve a single ordinary differential equation, given in equation \eqref{H11eom}.  The bulk viscosity to shear viscosity ratio can then be extracted from the solution of \eqref{H11eom} by using equation \eqref{GotZetaOverEta}, which is our main result.  In examples, a conjectured bound \cite{Buchel:2007mf} on the ratio $\zeta/\eta$ of bulk to shear viscosity is observed to be satisfied in most but not all circumstances.

The structure of the rest of this paper is as follows.  In section~\ref{ANALYTIC} we explain our method for computing the bulk viscosity from AdS/CFT\@.  Section~\ref{BACKGROUND} summarizes the relevant gravitational backgrounds.  Section~\ref{BULKPERTURB} shows how to perturb the backgrounds in a rotationally invariant fashion.  Section~\ref{SHEARPERTURB} reviews the treatment of metric perturbations that control the shear viscosity, explaining in particular how to extract the imaginary part of the retarded Green's function of an off-diagonal component $T_{12}$ of the gauge theory stress tensor.  Section~\ref{BULKONSHELL} adapts this treatment to the case of bulk viscosity, using equations derived in section~\ref{BULKPERTURB}.  Section~\ref{SOLVESHEAR} detours once again to a review of the shear viscosity computation, showing how to solve the linearized equation of motion in the small $\omega$ limit and giving the final prescription for extracting $\eta$.  Section~\ref{SOLVEBULK} explains the analogous computation for the bulk viscosity.
In section~\ref{EXAMPLES} we compute the bulk viscosity for specific backgrounds. For the Chamblin-Reall backgrounds we obtain an analytic result for $\zeta/\eta$ in section~\ref{CHAMBLINREALL}, where we also give a different derivation of the same result using a Kaluza-Klein reduction argument.
In section~\ref{NUMERICAL}, we show some results of numerical computations of the bulk viscosity in a particular class of backgrounds whose equation of state is similar to that of QCD\@.  For completeness, we provide derivations of the Kubo formulas relevant to our results in appendix~\ref{KUBO}. In appendix~\ref{GAUGE} we clear up a subtlety in the computation of two-point functions related to our choice of gauge, and in appendix~\ref{AppGR} we justify
the prescription for the computation of the retarded two-point function that we used  in section~\ref{ANALYTIC}.

\section{Computation of viscosity using AdS/CFT}
\label{ANALYTIC}

\subsection{The background geometry}
\label{BACKGROUND}

We consider backgrounds with a scalar coupled to gravity.  The relevant action is
 \eqn{StartingAction}{
  S = {1 \over 16\pi G_5} \int d^5 x \, \sqrt{-g} \left[
    R - {1 \over 2} (\partial\phi)^2 - V(\phi) \right] \,,
 }
and the corresponding equations of motion are
 \eqn{StartingEoms}{
  \square\phi &= V'(\phi) \qquad
  R_{\mu\nu} - {1 \over 2} R g_{\mu\nu} = \tau_{\mu\nu} \,,
 }
where the stress tensor for the scalar is
 \eqn{taumunuDef}{
  \tau_{\mu\nu} = {1 \over 2} \partial_\mu \phi \partial_\nu \phi -
    {1 \over 4} g_{\mu\nu} (\partial\phi)^2 -
    {1 \over 2} g_{\mu\nu} V(\phi) \,.
 }
The backgrounds of interest have the form
 \eqn{Backgrounds}{
  ds^2 = e^{2A(r)} \left[ -h(r) dt^2 + d\vec{x}^2 \right] +
    e^{2B(r)} {dr^2 \over h(r)} \qquad \Phi = \phi(r) \,.
 }
The choice of radial variable $r$ is arbitrary: reparameterizing it leads only to a different choice of $B$.  A convenient choice for the backgrounds we will study is $r=\Phi$, namely the unperturbed value of $\phi(r)$.  Plugging \eno{Backgrounds} into \eno{StartingEoms} leads, with this choice, to
 \eqn{BackgroundEoms}{
  A'' - A' B' + {1 \over 6} &= 0  \cr
  h'' + (4A'-B') h' &= 0  \cr
  6 A' h' + h (24A'^2-1) + 2 e^{2B} V &= 0  \cr
  4A' - B' + {h' \over h} - {e^{2B} \over h} V' &= 0 \,,
 }
where primes denote $d/d\Phi$.  Using a numbering scheme $(t,x^1,x^2,x^3,r) = (x^0,x^1,x^2,x^3,x^5)$, the first two of these equations come from the $00$ and $11$ Einstein equations; the third comes from the $55$ Einstein equation; and the last comes from the scalar equation of motion.  There is typically some redundancy in equations obtained from classical gravity, with or without matter.  In the case of \eno{BackgroundEoms}, the redundancy is that the $\Phi$ derivative of the third equation follows algebraically from the four equations listed.

\subsection{Rotationally invariant deformations}
\label{BULKPERTURB}

In order to extract transport coefficients, we must calculate the $\omega \to 0$ limit of two-point functions of the schematic form
 \eqn{SchematicCorrelator}{
  G(\omega) = \int dt \, d^3 x \, e^{i \omega t}
    \langle {\cal O}(t,\vec{x}) {\cal O}(0,0) \rangle \,.
 }
As we explain in more detail in appendix~\ref{KUBO}, when calculating the bulk viscosity, the operator ${\cal O}$ may be taken to be $T_i^{\phantom{i}i} = T_{11} + T_{22} + T_{33}$.  Because this combination preserves the $SO(3)$ symmetry of spatial rotations, and because we integrate in \eno{SchematicCorrelator} over spatial separations $\vec{x}$ in an $SO(3)$-symmetric fashion, we should be capable of framing the entire calculation of bulk viscosity in an $SO(3)$-invariant fashion.  Imposing this symmetry on the metric ansatz forces all components of $g_{\mu\nu}$ to depend only on $t$ and $\Phi$, and it also forces $g_{11}=g_{22}=g_{33}$ and the vanishing of $g_{i0}$, $g_{i5}$, and $g_{ij}$ for $i \neq j$.  Gauge freedom allows one to stipulate that $\Phi$ doesn't change at all, and also that $g_{05}=0$.\footnote{Stipulating that $\Phi$ doesn't change and $g_{05}=0$ is a convenient alternative to the more standard axial gauge, where $\Phi$ is allowed to change but $g_{05}=0$ and $g_{55}$ is unperturbed.  The reason that two perturbations may be set equal to zero is that $SO(3)$-invariant diffeomorphisms $x^\mu \to x^\mu + \xi^\mu$ are specified by two independent functions of $t$ and $\Phi$, namely $\xi^0$ and $\xi^5$. There is a subtlety related to our choice of gauge that we clarify in appendix~\ref{GAUGE}.}  To summarize, the perturbed metric takes the form
 \eqn{PerturbedMetric}{
  g_{\mu\nu} = \begin{pmatrix} g_{00} & 0 & 0 & 0 & 0  \cr
    0 & g_{11} & 0 & 0 & 0  \cr
    0 & 0 & g_{11} & 0 & 0  \cr
    0 & 0 & 0 & g_{11} & 0  \cr
    0 & 0 & 0 & 0 & g_{55} \end{pmatrix}
 }
where
 \eqn{gmunuExpress}{
  g_{00} &= -e^{2A} h
    \left( 1 + {\lambda \over 2} H_{00} \right)^2  \cr
  g_{11} &= e^{2A} \left( 1 + {\lambda \over 2} H_{11} \right)^2  \cr
  g_{55} &= {e^{2B} \over h} \left( 1 + {\lambda \over 2}
    H_{55} \right)^2 \,.
 }
$A$, $B$, and $h$ are functions of $\Phi$ only which satisfy the equations \eno{BackgroundEoms}; $H_{00}$, $H_{11}$, and $H_{55}$ depend on $t$ and $\Phi$; and $\lambda$ is a formal expansion parameter.  The reason for the precise form \eno{gmunuExpress} is that it leads to a relatively simple expression for $\sqrt{-g}$.

In the presence of the ansatz \eno{PerturbedMetric}, the non-trivial components of Einstein's equations are for $\mu\nu = 00$, $11$, $55$, and $15$.  The scalar equation of motion is also a non-trivial constraint on the metric components.  These five equations may be formulated in terms of $g_{00}$, $g_{11}$, and $g_{55}$, but their explicit form is complicated and would not add clarity to the current exposition.  Let us instead pass directly to the linearized approximation, where we keep terms only up to $O(\lambda)$ in the equations of motion.  Let us also assume harmonic time dependence by replacing $H_{\mu\nu}(t,\Phi) \to e^{-i\omega t} H_{\mu\nu}(\Phi)$.  Because of some redundancy in the five equations of motion, similar to the redundancy of \eno{BackgroundEoms}, one winds up with only three independent equations, namely
 \begin{subequations}
 \begin{eqnarray}
  H_{11}'' &=& \left( -{1 \over 3A'} - 4A' + 3B' -
     {h' \over h} \right) H_{11}' +
    \left( -{e^{-2A+2B} \over h^2} \omega^2 +
     {h' \over 6h A'} - {h' B' \over h} \right) H_{11}
     \label{H11eom} \\
  H_{00}' &=& {1 \over 12 h^2 A'^2} \Bigg[
   2h^2 \left( 1-6 A'^2-3A' {h' \over h} \right) H_{11}' \nonumber  \\
   &&\qquad\quad{} -
   \left( hh' (1-24A'^2) - 6 A' h'^2 - 12 A' e^{-2A+2B} \omega^2
     \right) H_{11} \Bigg]
     \label{H00eom} \\
  H_{55} &=& {1 \over A'} \left[ H_{11}' - {h' \over 2h} H_{11}
    \right] \,.
    \label{H55eom}
 \end{eqnarray}
 \end{subequations}
These equations are usually too complicated to solve in closed form; indeed, the same is true of the equations \eno{BackgroundEoms} that determine $A$, $B$, and $h$.  However, a combination of series expansions near the horizon (where $h$ has a simple zero as a function of $\Phi$) and numerics can be used to find fairly precise solutions.  The terms proportional to $\omega^2$ may be ignored except very close to the horizon.  The horizon boundary conditions appropriate for computing a retarded Green's function are that only an infalling wave is allowed at the horizon.

With a solution to \eqref{H11eom}--\eqref{H55eom} in hand that satisfies appropriate boundary conditions at the horizon, one still has to evaluate the on-shell action in order to find Green's functions, using the prescription of \cite{Gubser:1998bc, Witten:1998qj}.  In fact, there are two subtleties involved in this evaluation:
 \begin{itemize}
  \item The Green's functions of interest are complex, whereas the action is real.  The transport coefficients are read off from the imaginary part of the Green's functions, which is closely related to a  conserved Noether current in a complexification of the supergravity action.
  \item Boundary terms must be added to the action \eno{StartingAction} in order to obtain a Lagrangian which depends only on the perturbations $H_{\mu\nu}$ and their first derivatives \cite{Gibbons:1976ue}.  Additional boundary terms are often needed to cancel divergent parts of the resulting Green's functions.
 \end{itemize}

\subsection{Shear perturbations}
\label{SHEARPERTURB}

In order to illustrate and resolve the two subtleties mentioned at the end of the previous section, it helps to start by rehearsing the now-standard case of the shear viscosity.  The metric, including  perturbations, needed in order to study the shear viscosity is
 \eqn{ShearPerturb}{
  g_{\mu\nu} = \begin{pmatrix} -e^{2A} h & 0 & 0 & 0 & 0  \cr
    0 & e^{2A} & e^{2A} \lambda H_{12} & 0 & 0  \cr
    0 & e^{2A} \lambda H_{12} & e^{2A} & 0 & 0  \cr
    0 & 0 & 0 & e^{2A} & 0  \cr
    0 & 0 & 0 & 0 & e^{2B}/h \end{pmatrix} \,,
 }
where, as usual, $H_{12}$ is a function only of $t$ and $\Phi$, and we have persisted in using the gauge $r=\Phi$. This should be compared to \eqref{PerturbedMetric}--\eqref{gmunuExpress} in the bulk viscosity case. At the level of linearized perturbations of the equations of motion, or quadratic perturbations of the action, $H_{12}$ decouples from all other metric perturbations.  It is straightforward to show, again to linear order in $\lambda$, that the only non-trivial equation of motion in the presence of the ansatz \eno{ShearPerturb} is the $12$ Einstein equation, which reads
 \eqn{EinsteinOneTwo}{
  H_{12}'' + \left( 4 A' - B' + {h' \over h} \right) H_{12}' +
    {e^{-2A+2B} \over h^2} \omega^2 H_{12} = 0 \,,
 }
where we have again substituted $H_{12}(t,\Phi) \to e^{-i\omega t} H_{12}(\Phi)$.  The result \eno{EinsteinOneTwo} in the shear viscosity computation is parallel to \eqref{H11eom}--\eqref{H55eom} in the bulk viscosity computation.  Assuming that we have found a solution (at least an approximate one) to \eno{EinsteinOneTwo}, we must now confront the evaluation of the on-shell action to $O(\lambda^2)$.  The first difficulty is that the substitution $H_{12}(t,\Phi) \to e^{-i\omega t} H_{12}(\Phi)$ doesn't make sense outside the context of a linearized equation.  So let's instead keep $H_{12}$ as a general (real) function of $t$ and $\Phi$.  The next difficulty is appropriately fixing total derivative terms.  Starting from \eno{StartingAction}, without any improvements or boundary terms added, one finds
 \eqn{Lfixing}{
  S &= {1 \over 16\pi G_5} \int d^5 x \, {\cal L}  \cr
  {\cal L} &= \hat{\cal L} + \partial_t \hat{\cal L}^t +
    \partial_\Phi \hat{\cal L}^\Phi  \cr
  \hat{\cal L} &= {1 \over 2h} e^{2A+B} \dot{H}_{12}^2 -
    {h \over 2} e^{4A-B} H_{12}'^2  \cr
  \hat{\cal L}^t &= -{2 \over h} e^{2A+B} H_{12} \dot{H}_{12}  \cr
  \hat{\cal L}^\Phi &= 2 h e^{4A-B} H_{12} H_{12}' +
    h A' e^{4A-B} H_{12}^2 \,.
 }
We will refer to ${\cal L}$ as the ``unimproved'' lagrangian and to $\hat{\cal L}$ as the ``improved'' lagrangian.  The improvement terms are total derivatives $\partial_t \hat{\cal L}^t$ and $\partial_\Phi \hat{\cal L}^\Phi$, and they come in part from the well-known Gibbons-Hawking term involving the trace of the extrinsic curvature integrated over the boundary of spacetime.  The form of the improvement terms is largely fixed by demanding that $\hat{\cal L}$ should depend only on $H_{12}$ and its first derivatives.  But this requirement doesn't constrain the second term in $\hat{\cal L}^\Phi$ as shown in \eno{Lfixing}: indeed, any multiple of $H_{12}^2$ could be added to $\hat{\cal L}^\Phi$, and correspondingly one would wind up with terms proportional to $H_{12} H_{12}'$ and $H_{12}^2$ in the improved lagrangian.  The particular choice of $\hat{\cal L}^\Phi$ we made in \eno{Lfixing} obviously leads to the simplest form of $\hat{\cal L}$, but in the treatment of the bulk viscosity it is not so obvious how to choose $\hat{\cal L}^\Phi$.  Let us therefore consider a slightly generalized ``improved'' action and lagrangian:
 \eqn{Lgeneralized}{
  \hat{S} &= {1 \over 16 \pi G_5} \int d^5 x \, \hat{\cal L}  \cr
  \hat{\cal L} &= {1 \over 2h} e^{2A+B} \dot{H}_{12}^2 -
    {h \over 2} e^{4A-B} H_{12}'^2 + {1 \over 2} G' H_{12}^2 +
    G H_{12} H_{12}' \,.
 }
$\hat{\cal L}$ evidently differs from the improved lagrangian in \eno{Lfixing} by the total derivative $\partial_\Phi \left( {1 \over 2} G H_{12}^2 \right)$.  Here $G$ is an arbitrary, smooth, real function of $\Phi$.  The idea (modulo the issue of real versus complex perturbations) is that evaluating $\hat{S}$ by plugging in a solution to the linearized equation of motion \eno{EinsteinOneTwo} gives a two-point function of the operator dual to $H_{12}$, namely $T_{12}$.  But as long as $G$ is not fixed by some first-principles consideration, this Green's function will be ambiguous!  Fortunately, we will soon see that this ambiguity affects only the real part of the Green's function, and we want the imaginary part.  This is not to say that $G$ cannot be determined: it is closely related to counterterms in the prescription for computing Green's functions which are widely studied: see for example \cite{Balasubramanian:1999re, Bianchi:2001kw, Skenderis:2002wp}.  The point is that we do not need to determine it.

Having side-stepped the more subtle aspects of boundary terms, let us move on to the issue of obtaining a complex-valued Green's function from a real-valued action.  It helps to pass to the following real-valued lagrangian for a complex-valued field $h_{12} = h_{12}(\Phi)$:
 \eqn{LCdef}{
  \hat{\cal L}_{\bf C} \equiv
   {\omega^2 \over h} e^{2A+B} |h_{12}|^2 -
    h e^{4A-B} |h_{12}'|^2 + G' |h_{12}|^2 +
    G (h_{12} h^{*\prime}_{12} + h^*_{12} h_{12}') \,.
 }
$\hat{\cal L}_{\bf C}$ is constructed so that the equation for $h_{12}$ obtained from it is identical to the one obtained for $H_{12}$ from the improved lagrangian in \eno{Lgeneralized} with the assumption of a harmonic time dependence $e^{-i\omega t}$---that is, it is the $12$ Einstein equation, \eno{EinsteinOneTwo}.  One may re-express $\hat{\cal L}_{\bf C}$ as a total derivative plus terms that vanish upon use of the equations of motion:
 \eqn{LCpartial}{
  \hat{\cal L}_{\bf C} = \partial_\Phi J +
    h^*_{12} \left( {\partial \hat{\cal L}_{\bf C} \over
      \partial h^*_{12}} -
   {d \over d\Phi} {\partial \hat{\cal L}_{\bf C} \over
     \partial h^{*\prime}_{12}} \right)
 }
where
 \eqn{Jdef}{
  J = -h e^{4A-B} h^*_{12} h_{12}' + G |h_{12}|^2 \,.
 }
Up to a real normalization factor and the ambiguity in $G(\Phi)$, the limit of $J$ as one approaches the conformal boundary is the quantity to be identified as the retarded Green's function.  This claim essentially follows the discussion of \cite{Son:2002sd}, which was justified using a first-principles application of the prescription of \cite{Gubser:1998bc, Witten:1998qj} in a Schwinger-Keldysh formalism in \cite{Herzog:2002pc}.  In appendix~\ref{AppGR}, we provide a different
derivation of this result.  Because we only care about the imaginary part of the Green's function for purposes of computing transport coefficients, we define
 \eqn{FluxDef}{
  {\cal F} = -\Im J = {1 \over 2i} h e^{4A-B} (h^*_{12} h_{12}' -
    h^{*\prime}_{12} h_{12}) \,.
 }
${\cal F}$ coincides precisely with the conserved Noether charge associated with the $U(1)$ symmetry of \eno{LCdef} under a phase rotation $h_{12} \to e^{i\theta} h_{12}$.  It has the physical interpretation of the number flux of gravitons in the radial direction.\footnote{Indeed, calculations of imaginary parts of Green's functions using conserved number fluxes were commonplace before the prescription \cite{Gubser:1998bc, Witten:1998qj} was suggested: see for example \cite{Das:1996wn,Klebanov:1997kc,Gubser:1997cm,Gubser:1997se}.}  The sign of ${\cal F}$ was chosen so that ${\cal F}>0$ corresponds to particles falling into the black hole.  ${\cal F}$ is evidently the Wronskian of $h_{12}$ and $h^*_{12}$, multiplied by the factor required by Abel's identity to make it constant when the equation of motion \eno{EinsteinOneTwo} is obeyed.  As such, its form up to an overall multiplicative constant could have been guessed directly from the equation of motion \eno{EinsteinOneTwo} without detailed considerations regarding the on-shell action.

\subsection{The number flux for rotationally invariant perturbations}
\label{BULKONSHELL}

In analogy to \eno{FluxDef}, it is natural to guess that the imaginary part of the retarded Green's function for $T_i^{\phantom{i}i}$ is proportional to a flux proportional to the Wronskian of a complexified solution $h_{11}$ with its conjugate:
 \eqn{GuessFlux}{
  {\cal F} = -{i e^{4A-B} h \over 8A'^2}
    (h^*_{11} h_{11}' - h^{*\prime}_{11} h_{11}) \,.
 }
The $\Phi$-dependence of the prefactor can be deduced from Abel's identity applied to \eqref{H11eom}, but in our current approach, the overall constant factors in \eno{GuessFlux} must be fixed by a calculation of the on-shell action.  We describe such a calculation in the next paragraph.  The algebra gets a little complicated because there are three fields involved, namely $H_{00}$, $H_{11}$, and $H_{55}$.  However, every step is analogous to the foregoing discussion of the on-shell action in the shear viscosity calculation.

The action \eno{StartingAction} in the presence of the ansatz \eno{PerturbedMetric} takes the form
 \eqn{Lbulk}{
  S &= {1 \over 16\pi G_5} \int d^5 x \, {\cal L}  \cr
  {\cal L} &= \hat{\cal L} + \partial_t \hat{\cal L}^t +
    \partial_\Phi \hat{\cal L}^\Phi  \cr
  \hat{\cal L} &= {1 \over 2} \dot{\vec{H}}^T {\bf M}^{tt}
     \dot{\vec{H}} +
   {1 \over 2} \vec{H}'^T {\bf M}^{\Phi\Phi} \vec{H}' +
   {1 \over 2} \vec{H}^T {\bf M} \vec{H} +
   \vec{H}'^T {\bf M}^\Phi \vec{H} +
   \partial_\Phi \left( {1 \over 2} \vec{H}^T {\bf G} \vec{H} \right)
 }
where
 \begin{subequations}
 \begin{equation}
  \vec{H} = \begin{pmatrix} H_{00} \\ H_{11} \\ H_{55} \end{pmatrix}
   \qquad
  {\bf M}^{tt} = {3e^{2A+B} \over h} \begin{pmatrix}
    0 & 0 & 0 \\ 0 & 1 & 1/2 \\ 0 & 1/2 & 0 \end{pmatrix}
 \end{equation}
 \begin{equation}
  {\bf M}^{\Phi\Phi} = -3e^{4A-B} h \begin{pmatrix}
    0 & 1/2 & 0 \\ 1/2 & 1 & 0 \\ 0 & 0 & 0 \end{pmatrix}
 \end{equation}
 \begin{equation}
  {\bf M} = -{3 \over 2} e^{4A-B} \left[ -h (1-24A'^2) +
    6A' h' \right] \begin{pmatrix} 0 & 1/2 & 0 \\
     1/2 & 1 & 0 \\ 0 & 0 & 1/6 \end{pmatrix}
 \end{equation}
 \begin{equation}
  {\bf M}^\Phi = -{3 \over 4} e^{4A-B} \begin{pmatrix}
    0 & 6hA' & -2hA' \\ 6hA'+h' & 2 (6hA'+h') & -6hA'-h' \\
     0 & 0 & 0 \end{pmatrix} \label{MPhiDef} \,,
 \end{equation}
 \end{subequations}
and ${\bf G}$ is a symmetric matrix which may depend on $\Phi$ in an arbitrary but smooth way.  In principle, ${\bf G}$ should be determined by insisting that the full action has only first derivative terms; however, as in the shear viscosity case, we do not require the explicit form of ${\bf G}$.  Likewise, we do not need to know the explicit form of the improvement terms $\hat{\cal L}^t$ and $\hat{\cal L}^\Phi$.  The real-valued lagrangian analogous to \eno{LCdef} for complex-valued fields $\vec{h} = \vec{h}(\Phi)$ is
 \eqn{LCagain}{
  \hat{\cal L}_{\bf C} =
   \vec{h}^{*\prime T} {\bf m} \vec{h}' +
    \vec{h}^{*T} {\bf k} \vec{h} +
    \vec{h}^{*\prime T} {\bf b} \vec{h} +
    \vec{h}^{*T} {\bf b}^{*T} \vec{h}'
 }
where
 \eqn{mkbDef}{
  {\bf m} = {\bf M}^{\Phi\Phi} \qquad
   {\bf k} = \omega^2 {\bf M}^{tt} + {\bf M} + {\bf G}' \qquad
   {\bf b} = {\bf M}^\Phi + {\bf G} \,.
 }
Just as in \eno{LCpartial}, we may express $\hat{\cal L}_{\bf C}$ as a total derivative plus terms that vanish on-shell:
 \eqn{LCagainreduced}{
  \hat{\cal L}_{\bf C} =
   \partial_\Phi J + \vec{h}^{*T}
     \left( {\partial \hat{\cal L}_{\bf C} \over
       \partial \vec{h}^*} -
      {d \over d\Phi} {\partial \hat{\cal L}_{\bf C} \over
       \partial \vec{h}^{*\prime}} \right)
 }
where
 \eqn{Jagain}{
  J = \vec{h}^{*T} ({\bf m} \vec{h}' + {\bf b} \vec{h}) \,.
 }
The conserved Noether charge associated with the symmetry $\vec{h} \to e^{i\theta} \vec{h}$ is
 \eqn{FluxAgain}{
  {\cal F} = -\Im J = {i \over 2} \left[
    \vec{h}^{*T} ({\bf m} \vec{h}' + {\bf b} \vec{h}) -
    (\vec{h}^{*\prime T} + \vec{h}^{*T} {\bf b}^T) \vec{h} \right] \,.
 }
The form \eno{FluxAgain} appears to be distinct from \eno{GuessFlux}.  But one can see from \eqref{H11eom}--\eqref{H55eom} that all quantities appearing in ${\cal F}$ as defined in \eno{FluxAgain} can be eliminated in favor of $h_{11}'$ and $h_{11}$.  It is straightforward to check that the result of such an elimination is precisely the form quoted in \eno{GuessFlux}.  As in the case of the shear viscosity, ${\cal F}$ has the physical interpretation of a conserved number flux of particles falling into the horizon.  The particles in question are a mixture of metric and scalar perturbations, though as we have seen, a gauge choice is possible that sets the scalar perturbations to zero.

\subsection{Low-frequency limit of shear perturbations}
\label{SOLVESHEAR}

With the conserved number flux in hand, we should be able to extract the imaginary part of the retarded Green's function given a solution of the equations of motion.  As before, it is illustrative to start by reviewing the shear viscosity computation.  The linearized equation of motion \eno{EinsteinOneTwo} can be solved for $\omega=0$:
 \eqn{ExactZero}{
  \hbox{$\omega=0$ solution:}\qquad
   h_{12} = a_{12} + b_{12} \int_0^\Phi dr \, {e^{-4A+B} \over h} \,,
 }
where $a_{12}$ and $b_{12}$ are integration constants and we have assumed that $\Phi\to 0$ at the conformal boundary.  Because $h$ has a simple zero at the horizon, $\Phi=\Phi_H$, the second term diverges logarithmically there.  In the strict limit $\omega=0$, this term is disallowed.  For very small $\omega$, we may use a matching procedure which is analogous to boundary layer theory and well known in the literature on absorption by black holes: see for example \cite{Das:1996we}.  The trick is to find the leading behavior of solutions for non-zero $\omega$ just outside the horizon:
 \eqn{NearHorizonH12}{
  \hbox{$\Phi\approx\Phi_H$ solution:}\qquad
   h_{12} = c_{12}^+ (\Phi_H-\Phi)^{i\omega/4\pi T} +
     c_{12}^- (\Phi_H-\Phi)^{-i\omega/4\pi T} \,,
 }
where the Hawking temperature is
 \eqn{THawking}{
  T = {-h'(\Phi_H) \over 4\pi} e^{A(\Phi_H)-B(\Phi_H)} \,,
 }
and we have assumed that $\Phi_H>0$ and $h \to 1$ as $\Phi\to 0$, i.e.~at the conformal boundary.  Standard horizon boundary conditions, associated with the enforcement of causality, are to suppress the solution corresponding to gravitons coming out of the horizon: that is, $c_{12}^+=0$.  The matching procedure is to expand \eno{ExactZero} around the horizon and \eno{THawking} around $\omega=0$ and extract relations among $a_{12}$, $b_{12}$, and $c_{12}^-$ by comparing the expansions.  Briefly, the result is
 \eqn{DoubleExpand}{
  h_{12} \approx a_{12} + b_{12} {e^{-4A(\Phi_H)+B(\Phi_H)} \over
    h'(\Phi_H)} \log(\Phi_H-\Phi) \approx
   c_{12}^- \left[ 1 - {i\omega \over 4\pi T} \log(\Phi_H-\Phi)
     \right]\,.
 }
From \eno{DoubleExpand} we find, to leading order in small $\omega$,
 \eqn{GotABC}{
  a_{12} = c_{12}^- \qquad
  b_{12} = i\omega e^{3A(\Phi_H)} c_{12}^- \,.
 }
The flux is most easily computed in the limit $\Phi \to \Phi_H$ from \eno{NearHorizonH12}:
 \eqn{FluxH}{
  {\cal F} = \omega e^{3A(\Phi_H)} |c_{12}^-|^2 \,,
 }
still at leading order in small $\omega$.

The imaginary part of the retarded Green's function is
 \eqn{RetardedGreens}{
  \Im G_R(\omega) = -{{\cal F} \over 16\pi G_5}
 }
where
 \eqn{DefineGreensH12}{
  G_R(\omega) = -i \int dt \, d^3 x \, e^{i\omega t}
    \theta(t) \langle [ T_{12}(t,\vec{x}), T_{12}(0,0) ] \rangle \,,
 }
and ${\cal F}$ is computed assuming $a_{12}=1$.  The shear viscosity is
 \eqn{ExtractShear}{
  \eta = -\lim_{\omega\to 0} {1 \over \omega} \Im G_R(\omega)
    = {e^{3A(\Phi_H)} \over 16 \pi G_5} \,.
 }
One may confirm the relation $\eta/s = 1/4\pi$ at this point simply by noting that the entropy density is given by $s = e^{3A(\Phi_H)}/4 G_5$.  It is also worth pointing out that we never needed the explicit result for $b_{12}$ quoted in \eno{GotABC}, because we computed the flux ${\cal F}$ at the horizon, not at the conformal boundary.  In fact, we only needed the $\omega = 0$ solution which is regular at the horizon and approaches one at the boundary;  $c_{12}^-$ is then given exactly by the value of $H_{12}$ at the horizon.

\subsection{Low frequency limit of rotationally invariant perturbations}
\label{SOLVEBULK}

The imaginary part of the retarded correlator of $T_i^{\phantom{i}i}$ can be obtained in a similar manner by solving \eqref{H11eom}.  The $\omega = 0$ equation doesn't seem to have an analytic solution in general, so it is usually necessary to proceed numerically.  The boundary conditions that we need to impose are $h_{11} (0) = 1$ and the fact that $h_{11}$ should be regular at $\Phi = \Phi_H$.  As in the shear viscosity case, this solution needs to be matched onto the boundary layer at $\Phi \approx \Phi_H$.

Close to the horizon, the two solutions of \eqref{H11eom} behave as
 \eqn{NearHorizonH11}{
  \hbox{$\Phi\approx\Phi_H$ solution:}\qquad
   h_{11} = c_{11}^+ (\Phi_H-\Phi)^{i\omega/4\pi T} +
     c_{11}^- (\Phi_H-\Phi)^{-i\omega/4\pi T} \,,
 }
where $T$ is the Hawking temperature as in \eqref{THawking}.  Infalling boundary conditions mean setting $c_{11}^+ = 0$.  For small $\omega$,
 \eqn{SmallOmegaNH}{
  h_{11} \approx c_{11}^- \left[1 - {i \omega\over 4 \pi T} \log(\Phi_H - \Phi) \right]\,,
 }
so to leading order in $\omega$ we have $h_{11} \approx c_{11}^-$.  We can therefore find $c_{11}^-$ by just evaluating the $\omega = 0$ solution at the horizon.

With the value of $c_{11}^-$ in hand, we can find $\cal F$ by plugging \eqref{NearHorizonH11} into \eqref{GuessFlux} and keeping only the leading order term in $\Phi_H - \Phi$.  The subleading terms do not give corrections to ${\cal F}$, because ${\cal F}$ is conserved so it should be independent of $\Phi$.  The fact that upon plugging \eqref{NearHorizonH11} into \eqref{GuessFlux} we also get terms suppressed by various powers of $\Phi_H - \Phi$ just shows that the asymptotic form \eqref{NearHorizonH11} is not exact and should in fact be corrected in such a way that the resulting ${\cal F}$ stays independent of $\Phi$.  We obtain
 \eqn{GotF}{
  {\cal F} = \omega e^{3 A(\Phi_H)} {|c_{11}^-|^2 \over 4 A'(\Phi_H)^2} \,.
 }

Given ${\cal F}$, we can extract the imaginary part of the retarded Green's function of $T_i^{\phantom{i}i}$ through\footnote{The coupling of the metric to the stress tensor is ${1 \over 2} H_{mn} T^{mn}$.  So when we set $H_{11}=H_{22}=H_{33}$, the dual operator is ${1 \over 2} T_i^{\phantom{i}i}$.  But when we set $H_{12}=H_{21}$, the operator we couple to is $T_{12}$.  That is why there are explicit factors of $1 \over 2$ in \eno{DefineGreensH11} but not in \eno{DefineGreensH12}.}
 \eqn{RetardedGreens2}{
  \Im G_R(\omega) = -{{\cal F} \over 16\pi G_5}\,,
 }
where
 \eqn{DefineGreensH11}{
  G_R(\omega) = -i \int dt \, d^3 x \, e^{i\omega t}
    \theta(t) \langle [
      \textstyle{1 \over 2} T_i^{\phantom{i}i}(t,\vec{x}),
      \textstyle{1 \over 2} T_k^{\phantom{k}k}(0,0) ] \rangle \,.
 }
The bulk viscosity is then given by the Kubo formula (see appendix~\ref{KUBO})
 \eqn{KuboBulk}{
  \zeta = -{4 \over 9} \lim_{\omega\to 0} {1 \over \omega} \Im G_R(\omega)
    = {s \over 36 \pi A'(\Phi_H)^2} |c_{11}^-|^2 \,,
 }
where we have used $s = e^{3 A(\Phi_H)}/4 G_5$.  Since Einstein's equations \eqref{BackgroundEoms} for the background imply $A'(\Phi_H) = - V(\Phi_H)/3 V'(\Phi_H)$, and $\eta/s = 1/4 \pi$, equation \eqref{KuboBulk} yields
 \eqn{GotZetaOverEta}{
  {\zeta\over \eta} = {V'(\Phi_H)^2 \over V(\Phi_H)^2} |c_{11}^-|^2 \,.
 }

In \cite{Gubser:2008ny} it is noted that the speed of sound in the background \eqref{Backgrounds} can be approximately expressed as
 \eqn{csapprox}{
  c_s^2 \simeq {1\over 3} - {1\over 2}{V'(\Phi_H)^2 \over V(\Phi_H)^2} \,.
 }
Combining this with \eqref{GotZetaOverEta} we obtain
 \eqn{ZetaOverEtaApprox}{
  {\zeta\over \eta} \simeq 2\left( {1\over 3} - c_s^2 \right) |c_{11}^-|^2 \,.
 }
It is interesting to compare this expression with the conjectured bound on $\zeta/\eta$ of \cite{Buchel:2007mf} stating that all $3+1$-dimensional finite-temperature field theories should satisfy
 \eqn{BuchelBound}{
  {\zeta\over \eta} \geq 2\left( {1\over 3} - c_s^2 \right) \,.
 }
We will refer to this inequality as the viscosity ratio bound.  Using \eqref{ZetaOverEtaApprox} we note that \eqref{BuchelBound} is roughly equivalent to $|c_{11}^-|^2 \geq 1$ --- roughly because \eqref{ZetaOverEtaApprox} itself is an approximate result.

\section{Examples}
\label{EXAMPLES}

\subsection{Chamblin-Reall backgrounds}
\label{CHAMBLINREALL}

The simplest application of the result \eno{GotZetaOverEta} is to Chamblin-Reall backgrounds \cite{Chamblin:1999ya}, which are gravity solutions following from the action \eqref{StartingAction} with potential
 \eqn{VCR}{
  V(\phi) = V_0 e^{\gamma \phi} \,,
 }
where $V_0<0$ and $0<\gamma<\sqrt{2/3}$ are constants.  In a gauge where $r = \Phi$, the coefficients of the Chamblin-Reall metric take the form
 \eqn{CRMetric}{
  e^{2 A} &= \exp \left(-{2\Phi \over 3 \gamma} \right) \qquad
   e^{2 B} = -{8 - 3\gamma^2\over 6 \gamma^2 V_0} \exp (-\gamma \Phi) \cr
  h &= 1 - \exp \left[-{8 - 3 \gamma^2 \over 6 \gamma} (\Phi_H - \Phi) \right]\,.
 }

An interesting feature of the Chamblin-Reall solutions is that the entropy density has a power-like dependence on temperature,
 \eqn{sCR}{
  s \propto T^{6/(2-3 \gamma^2)} \,.
 }
If this background were dual to a field theory, equation \eqref{sCR} would imply that the speed of sound in such a theory is constant and given by
 \eqn{csCR}{
  c_s^2 \equiv {d \log T \over d \log s} = {1\over 3} - {\gamma^2\over 2} \,.
 }
The first equality above is true in any field theory at zero chemical potential.

It is straightforward to check that for a Chamblin-Reall background the coefficient of $h_{11}$ in \eqref{H11eom} vanishes when $\omega = 0$.  Consequently, $h_{11} = \textrm{const}$ is a solution to the $\omega = 0$ equation.  Since the other linearly independent solution diverges  logarithmically at the horizon, and since one requires $h_{11} \to 1$ in the ultraviolet, one concludes that $c_{11}^- \equiv h_{11}(\Phi_H) = 1$ in this case.  The ratio of bulk to shear viscosity can then be easily found from \eqref{GotZetaOverEta} to be
 \eqn{GotZetaCR}{
  {\zeta \over \eta} = \gamma^2 = 2 \left( {1\over 3} - c_s^2 \right) \,.
 }
This shows that, as was already pointed out in \cite{Buchel:2007mf}, the
Chamblin-Reall backgrounds saturate the viscosity ratio bound \eqref{BuchelBound}.

There is a more conceptual way of understanding \eqref{GotZetaCR} that comes from the fact that the Chamblin-Reall metric \eqref{CRMetric} can be obtained through a Kaluza-Klein reduction from a $D+1$-dimensional anti-de Sitter space $AdS_{D+1}$.  (Related arguments are presented in \cite{Mas:2007ng,Buchel:2007mf}.)  The Kaluza-Klein reduction is explained in \cite{Gubser:2008ny}, but the details of this reduction will not be relevant in the following argument.  All we will need  is the relation between $\gamma$ and $D$:
 \eqn{GotD}{
  D = {8 - 3 \gamma^2 \over 2 - 3 \gamma^2}  \,.
 }
In any field theory in $d$ spacetime dimensions, the bulk and shear viscosities are related to the two-point functions of the stress-energy tensor by the generalization of the Kubo formula \eqref{Kubo}:
 \eqn{Kubod}{
  \eta \left(\delta_{ik} \delta_{jl} + \delta_{il} \delta_{jk} - {2\over d-1} \delta_{ij} \delta_{kl} \right) + \zeta \delta_{ij} \delta_{kl} =
   \lim_{\omega\to 0} {1\over \omega} \int d^3 x\, dt\, e^{i \omega t} \theta(t)
   \langle [ T_{ij} (t, \vec{x}), T_{kl} (0, 0) ] \rangle \,,
 }
where all indices denote spatial components.  Setting $i = j$ and $k = l$ and summing over $i$ and $k$ running from $1$ to $3$, we obtain
 \eqn{CRKubo}{
  9 \zeta_{\rm CR} = \lim_{\omega\to 0} {1\over \omega} \int d^3 x\, dt\, e^{i \omega t} \theta(t)
   \sum_{i, k = 1}^3 \langle [ T_{ii} (t, \vec{x}), T_{kk} (0, 0) ] \rangle_{\rm CR}
 }
in the Chamblin-Reall case and
 \eqn{CFTKubo}{
  \eta_{\rm CFT} \left( 6 - {18 \over {D-1}} \right)
   = \lim_{\omega\to 0} {1\over \omega} \int d^3 x\, dt\, e^{i \omega t} \theta(t)
   \sum_{i, k = 1}^3 \langle [ T_{ii} (t, \vec{x}), T_{kk} (0, 0) ] \rangle_{\rm CFT} \,,
 }
in the ${\rm CFT}_D$ dual to the anti-de Sitter space $AdS_{D+1}$ from which the Chamblin-Reall solution was obtained.  In \eqref{CFTKubo} we used the fact that $\zeta_{\rm CFT} = 0$ by scale invariance, so there is no bulk viscosity contribution.

Because field theory two-point functions can be computed from taking functional derivatives of the on-shell gravitational action, the two-point function in the right-hand side of \eqref{CRKubo} differs from the one in \eqref{CFTKubo} by the ratio $G_{D+1}/G_5$ of the Newton constants in $D+1$ and in $5$ dimensions.  Consequently,
 \eqn{TwoPt}{
  {\sum_{i, k = 1}^3 \langle [ T_{ii} (t, \vec{x}), T_{kk} (0, 0) ] \rangle_{\rm CR} \over \sum_{i, k = 1}^3 \langle [ T_{ii} (t, \vec{x}), T_{kk} (0, 0) ] \rangle_{\rm CFT}} = {G_{D+1} \over G_5} = {s_{\rm CR} \over s_{\rm CFT}} = {\eta_{\rm CR} \over \eta_{\rm CFT}}\,.
 }
In the last equality we used the fact that $\eta/s = 1/4 \pi$ in both theories.  Combining \eqref{TwoPt} with \eqref{CRKubo} and \eqref{CFTKubo}, it follows that
 \eqn{GotZetaCRAgain}{
  {\zeta_{\rm CR} \over \eta_{\rm CR}} = 2 \left({1\over 3} - {1\over D-1} \right) \,.
 }
Since \eqref{csCR} and \eqref{GotD} imply that in the Chamblin-Reall background $c_s^2 = 1/(D-1)$, one can immediately recover \eqref{GotZetaCR}.  If we now set $D=6$ we obtain $\zeta/\eta = 4/15$. This result was obtained in \cite{Benincasa:2006ei} through different methods.

\subsection{Numerical results on QCD-like models}
\label{NUMERICAL}

With \eqref{GotZetaOverEta} in hand, we can compute $\zeta/s$ numerically for any choice of the scalar potential $V(\phi)$ over a range of temperatures.  In this section we will present results for a specific form of $V(\phi)$ chosen so that the equation of state of the black hole resembles that of QCD \cite{Gubser:2008ny,Gubser:2008yx}:
 \eqn{QCDLikePotential}{
  V(\phi) = - {12\over L^2} \cosh(\gamma \phi) + b \phi^2 \,.
 }
Following the nomenclature of \cite{Gubser:2008yx}, we describe black holes derived by coupling a scalar with this potential to gravity as Type~I black holes.  The quantity $L$ is the radius of curvature of the asymptotic AdS geometry. We choose to work in units where $L=1$.  We choose $\gamma \approx 0.606$ in \eno{QCDLikePotential} in order for the squared speed of sound in the IR to be approximately $c_s^2 \approx 0.15$, a value broadly consistent with hadron gas phenomenology.  Similarly, the parameter $b$ appearing in \eqref{QCDLikePotential} can be adjusted so that the dimension $\Delta$ of the field theory operator dual to the bulk field $\phi$ matches that of $\tr F_{\mu\nu}^2$ in QCD at a particular scale.  As explained in \cite{Gubser:2008yx}, $\Delta \approx 3.93$ at an energy scale $Q = 3\, {\rm GeV}$, which gives $b \approx 2.057$.  If the matching is done at $Q = 1\, {\rm GeV}$, one obtains $\Delta \approx 3.61$ and $b \approx 1.503$.
\begin{figure}
	\centering
		\includegraphics{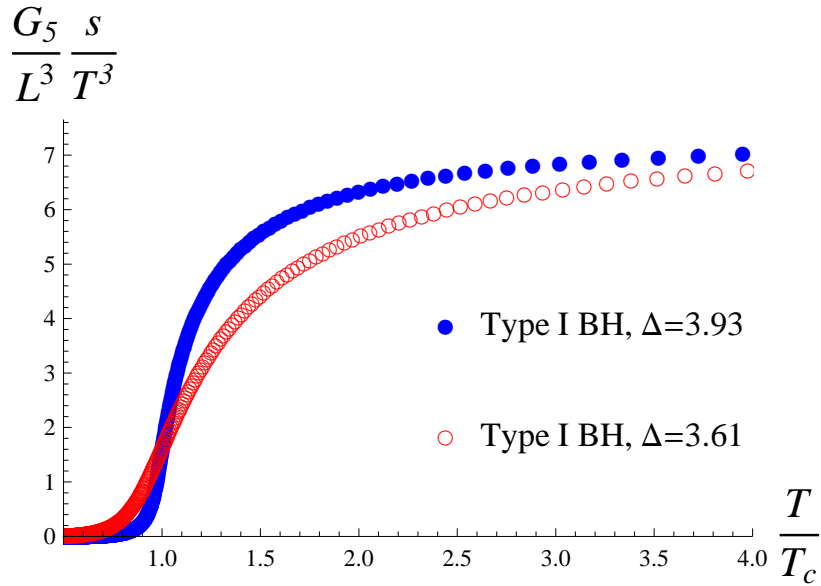}
	\caption{$s/T^3$ as a function of $T/T_c$ for two potentials of the form \eqref{QCDLikePotential} with
	\mbox{$\{\gamma\approx 0.606,b\approx 2.06,\Delta\approx 3.93\}$}
	and $\{\gamma\approx 0.606,b \approx 1.503,\Delta \approx 3.61\}$.  For each curve, $T_c$ is defined to be the ordinate of its inflection point.  These potentials were obtained in \cite{Gubser:2008ny,Gubser:2008yx}
	where their equations of state are shown to mimic that of QCD.}
	\label{fig:QCDLikePotentialEOS}
\end{figure}
Figures~\ref{fig:QCDLikePotentialEOS} and~\ref{fig:QCDLikePotentialZeta} show the equation of state and $\zeta/s$ corresponding to \eqref{QCDLikePotential} for these two choices of $b$.  Figure~\ref{fig:QCDLikePotentialZeta} should be compared to figure~1 of \cite{Karsch:2007jc}, representing the ratio $\zeta/s$ for QCD as computed from lattice data.  Similarly to what was found in \cite{Karsch:2007jc}, $\zeta/s$ does increase as $T$ approaches $T_c$ from above.  However, this quantity rises only up to $\left.\zeta/s\right|_\textrm{max} \approx 0.06$ around $T_c$, in contrast with the sharp rise exhibited in \cite{Karsch:2007jc}.

It is worth mentioning that the values of $\zeta/s$ obtained in these examples violate the viscosity ratio bound \eqref{BuchelBound} for a range of temperatures below $T_c$.  As can be seen from figure~\ref{fig:QCDLikePotentialZeta}, in the case $\Delta \approx 3.93$ the bound is violated for $0.541 \lesssim T/T_c \lesssim 0.900$, while in the case $\Delta \approx 3.61$ it is violated for $0.345 \lesssim T/T_c \lesssim 0.737$.  Even though the violations are smaller than $5\%$ for the potential with $\Delta \approx 3.93$ and smaller than $3\%$ for the one with $\Delta \approx 3.61$,  they are well within the margin of numerical error.  But although the potential \eno{QCDLikePotential} is similar to those encountered in gauged supergravity, its precise form was made up.  So the possibility remains that the bound \eno{BuchelBound} could be true in all potentials actually derivable from string theory.

\begin{figure}
	\centerline{\includegraphics[width=7in]{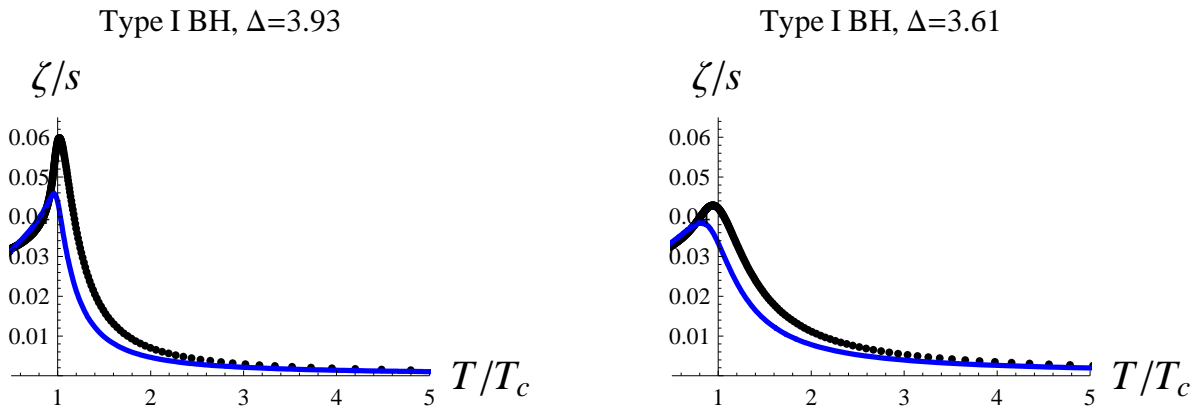}}
	\caption{The ratio $\zeta/s$ (dotted) for two potentials of the form \eqref{QCDLikePotential}, whose equations of state were plotted in figure~\ref{fig:QCDLikePotentialEOS}.  The solid curves show the Buchel bound \eqref{BuchelBound}, and a violation is evident when $T$ is below $T_c$ for
     both potentials.
	}
	\label{fig:QCDLikePotentialZeta}
\end{figure}

By solving \eqref{H11eom} for finite values of $\omega$ and computing the flux ${\cal F}$ as given in \eqref{GuessFlux}, we can also evaluate the spectral
density function given by
\eqn{Spectral}{
\rho(\omega) \equiv - {1\over \pi} \Im G_R(\omega) = {{\cal F} \over 16 \pi^2 G_5} \,.
}
As is explained in \cite{Karsch:2007jc}, in QCD $\rho \sim \omega^4$ for large values of $\omega$.  In our model,
preliminary numerical studies at large $\omega$ suggest a power-law behavior $ \Im G_R(\omega) \propto \omega^\nu$ but with
$\nu$ depending on temperature.  For instance, for the potential of the form
\eqref{QCDLikePotential} with $\{\gamma\approx 0.606,b = 2.06,\Delta \approx 3.93\}$ we found that
$\nu \approx 2.8$ for $T/T_c \approx 1.05$ and that $\nu \approx 3.2$  for $T/T_c \approx 1.45$. These values of $\nu$ were obtained from power-law fits in the range $5 \leq \omega/T \leq 20$, which may be too small a window to accurately capture the high-$\omega$ behavior.

\section*{Acknowledgments}

We thank D.~Kharzeev and H.~Meyer for useful discussions and especially A.~Nellore for collaboration on related projects.  This work was supported in part by the Department of Energy under Grant No.\ DE-FG02-91ER40671 and by the NSF under award number PHY-0652782.  F.D.R.~was also supported in part by FCT grant SFRH/BD/30374/2006.

\clearpage

\appendix

\section{Kubo formulas for bulk and shear viscosity}
\label{KUBO}

For completeness, we now outline the proof of the general Kubo formula for shear and bulk viscosity
 \eqn[c]{Kubo}{
  \eta \left(\delta_{ik} \delta_{jl} + \delta_{il} \delta_{jk} - {2\over 3} \delta_{ij} \delta_{kl} \right) + \zeta \delta_{ij} \delta_{kl} =
   -\lim_{\omega \to 0} {1\over \omega} \Im G^R_{ij, kl}(\omega) \cr
  G^R_{ij, kl}(\omega) \equiv -i \int d^3 x\, dt\, e^{i \omega t} \theta(t)
   \langle [ T_{ij} (t, \vec{x}), T_{kl} (0, 0) ] \rangle\,,
 }
from which one can obtain \eqref{ExtractShear} by setting $i = k = 1$ and $j = l = 2$, and equation \eqref{KuboBulk} by tracing over $ij$ and $kl$.  All indices in \eqref{Kubo} run over spatial directions only. The proof consists of two parts.  In section~\ref{LINEARRESPONSE} we apply Kubo's linear response theory to express the stress-energy response to metric perturbations in terms of the retarded two-point functions of $T_{ij}$.  In section~\ref{HYDRO} we extract the shear and bulk viscosities by comparing the results in section~\ref{LINEARRESPONSE} to linearized hydrodynamics.  The results included in this section are not new:  Kubo's linear response theory was developed about fifty years ago in \cite{Kubo1957, Kubo1966}, and the stress-energy response to metric fluctuations in linearized hydrodynamics is explained, for example, in \cite{Son:2007vk}.

\subsection{Kubo's linear response theory}
\label{LINEARRESPONSE}

We start with a finite-temperature quantum mechanical system whose Hamiltonian is
 \eqn{HPert}{
  H = H_0 + H' \qquad H' = -\lambda {\cal O} \delta(t) \,,
 }
where ${\cal O}$ is some time-independent Schr\"odinger operator and $\lambda$ is a small parameter.  In units where $\hbar = 1$, the equation of motion for the density matrix,
 \eqn{rhoEOM}{
  i {\partial \rho \over \partial t} = [H, \rho]\,,
 }
can be solved through time-dependent perturbation theory:
 \eqn{Gotrho}{
  \rho(t) = e^{-i H_0 t} \rho_0 e^{i H_0 t}
   + \delta \rho(t) \qquad \delta \rho(t)
    = i e^{-i H_0 t} [\lambda {\cal O}, \rho_0] e^{i H_0 t} \theta(t)
 }
where $\rho_0$ is the unperturbed density matrix.  Note that the above equations are all written in the Schr\"odinger picture where $\rho$ evolves in time, but ${\cal O}$ doesn't.

The expectation value $\langle {\cal P}(t) \rangle$ of a different operator ${\cal P}$ at a later time $t$ can then be computed from
 \eqn{GotP}{
  \langle {\cal P}(t) \rangle &\equiv \tr \left\{ {\cal P} \rho(t) \right\}
   = \tr \left\{{\cal P} e^{-i H_0 t} \rho e^{i H_0 t}\right\}
   + \delta \langle {\cal P}(t) \rangle \cr
  \delta \langle {\cal P}(t) \rangle
   &= i \tr \left\{ e^{-i H_0 t} [\lambda {\cal O}, \rho_0]
   e^{i H_0 t} {\cal P} \right\}\,.
 }
Using the cyclic property of the trace and passing to the Heisenberg picture, one obtains
 \eqn{GotDeltaP}{
  \delta \langle {\cal P}(t) \rangle = i \theta(t) \tr \left\{ \rho [{\cal P}(t), \lambda {\cal O}(0)] \right\} = i  \theta(t) \langle [{\cal P}(t), \lambda {\cal O} (0)] \rangle \,.
 }
The factor of $\theta(t)$ in the above formula can be understood from the fact that the perturbation, which occurs at $t=0$, can only affect the expectation value of ${\cal P}$ at later times.  Similarly, the commutator between ${\cal P}(t)$ and ${\cal O}(0)$ arises because it is a measure of the degree to which the perturbation can affect ${\cal P}(t)$.  In particular, if ${\cal P}(t)$ and ${\cal O}(0)$ commute, then the expectation value of ${\cal P}$ at time $t$ is unaffected by the perturbation, so $\langle \delta {\cal P}(t) \rangle=0$.

The result \eqref{GotDeltaP} can be applied to a field theory on ${\bf R}^{3, 1}$ whose lagrangian density is given by
 \eqn{LPert}{
  {\cal L} = {\cal L}_0 + {1\over 2} h_{kl} T^{kl} \,,
 }
where ${\cal L}_0$ is the unperturbed lagrangian density, $h_{kl}$ denotes perturbations of the metric, and $T^{kl}$ is the stress-energy tensor.  Again, indices $k$ and $l$ run only over spatial directions.  Taking $h_{kl}(t, \vec{x}) =  h_{kl} \delta^3 (\vec{x}) \delta(t)$, and using \eqref{GotDeltaP} with $\lambda {\cal O} = h_{kl} \delta^3(\vec{x})$ and ${\cal P}(t) = T_{ij}(t, \vec{x})$, one obtains
 \eqn{TklPert}{
  \delta \langle T_{ij}(t, \vec{x}) \rangle
   = {i\over 2} \theta(t) \langle [T_{ij} (t, \vec{x}), h_{kl} T^{kl} (t, 0)] \rangle \,.
 }
Because small perturbations add linearly, we can extend \eqref{TklPert} from perturbations localized at $\vec{x} = 0$ and $t = 0$ to spatially homogeneous perturbations where $h_{ij}(t, \vec{x}) = h_{ij}(t)$:
 \eqn{TklPertAgain}{
  \delta \langle T_{ij}(t, \vec{x}) \rangle
   = {i\over 2} \int dt' d^3 x' \, \theta(t-t') \langle[T_{ij}(t, \vec{x}), T^{kl}(t', \vec{x}')] \rangle h_{kl}(t')\,.
 }
In Fourier space, this equation reads
 \eqn{TklFourier}{
  \delta \langle T_{ij}(\omega, \vec{x}) \rangle
   = -{1\over 2} h_{kl}(\omega) G_{ij}^{R, kl}(\omega) \,,
 }
where $G_{ij, kl}^R(\omega)$ was defined in \eqref{Kubo}.

\subsection{Spatially homogeneous perturbations in linearized hydrodynamics}
\label{HYDRO}

In a curved background, the stress-energy tensor of a viscous fluid with velocity field $u^\mu$, energy density $\epsilon$, pressure $p$, and bulk and shear viscosities $\zeta$ and $\eta$ is given by
 \eqn{ViscFluid}{
  T^{\mu\nu} = (\epsilon + p) u^\mu u^\nu + p g^{\mu\nu}
   - P^{\mu\alpha} P^{\nu\beta} \left[ \eta \left(\nabla_\alpha u_\beta
   + \nabla_\beta u_\alpha - {2\over 3} g_{\alpha \beta} \nabla_\lambda u^\lambda \right)
   + \zeta g_{\alpha \beta} \nabla_\lambda u^\lambda \right] \,,
 }
where $P^{\mu\nu} \equiv g^{\mu\nu} + u^\mu u^\nu$ is a projection operator and $\nabla_\mu$ denotes covariant differentiation.  We are only interested in perturbing a fluid at rest on ${\bf R}^3$ by changing the space-space components of the metric to $g_{ij} = \eta_{ij} + h_{ij}$.  In this case, the velocity field $u^\mu = (1, 0, 0, 0)$ doesn't change to first order in $h_{ij}$ because of parity symmetry.  It can be checked that to lowest non-vanishing order in the perturbation, the change in the space-space components of the stress-energy tensor is given by
 \eqn{TklHydro}{
  \delta T_{ij} = p h_{ij} - {1\over 2} K  h_k^{\phantom{k} k} \delta_{ij}+ \delta_{ij} \eta \left(- \partial_t h_{ij}
   + {1\over 3} \delta_{ij} \delta^{kl} \partial_t h_{kl} \right)
   - {1\over 2} \zeta \delta_{ij} \delta^{kl} \partial_t h_{kl} \,,
 }
where $K \equiv -V \partial p / \partial V$ is the bulk modulus. In frequency space, this relation reads
  \eqn{TklHydroFT}{
  \delta T_{ij} = h_{kl}(\omega) \left(
  p \delta_i^{\phantom{i}k} \delta_j^{\phantom{j}l}
  - {1\over 2} K \delta_{ij} \delta^{kl}
  \right)
   + {i \omega \over 2} h_{kl}(\omega) \left[\eta \left(\delta_i^{\phantom{i}k} \delta_j^{\phantom{j}l}
   + \delta_i^{\phantom{i}l} \delta_j^{\phantom{j}k} -{2\over 3} \delta_{ij} \delta^{kl} \right)
   + \zeta \delta_{ij} \delta^{kl} \right] \,.
 }
Because linearized hydrodynamics is expected to be valid in the limit of slowly-varying perturbations, one should compare this formula to equation \eqref{TklFourier}.  Equation \eqref{Kubo} follows immediately.

\section{Gauge transformations and dual operators}
\label{GAUGE}

It is standard in AdS/CFT to assert that the leading behavior of the $H_{mn}$ component of the metric perturbations translates into a deformation of the field theory lagrangian by the operator ${1\over 2} T_{mn}$, provided that one works in ``axial gauge'' where $H_{5m} = 0$.  But we work in a gauge where $\delta\phi=0$, and as a result $H_{55}$ is not $0$.  The goal of this section is to justify the fact that, despite this gauge choice, the operator dual to $H_{11}=H_{22}=H_{33}$ is still given by ${1\over 2} T_i^{\phantom{i}i}$, as would be the case in axial gauge.  Most of the discussion will rely on the following asymptotic behaviors of various metric components near the boundary of AdS (see \cite{Gubser:2008ny}), written in a gauge where $r = \Phi$:
 \eqn[c]{Asymp}{
  A \approx {1\over {\Delta - 4}} \log \Phi  \qquad
   B \approx - \log \Phi \qquad
   h \approx 1 \cr
  \delta \phi \approx \alpha \Phi \qquad
   H_{mn} \approx R_{mn} \,.
 }
Here $R_{mn}$ and $\alpha$ are constants, and $\Delta$ is the dimension of the operator ${\cal O}_\phi$ dual to $\phi$.

Under an infinitesimal gauge transformation, the metric and scalar perturbations transform as follows:
 \eqn{Gauge}{
  \delta g_{mn} &\to \widetilde{\delta g}_{mn} = \delta g_{mn} + \nabla_m \xi_n + \nabla_n \xi_m \cr
  \delta \phi &\to \widetilde{\delta \phi} = \delta \phi + \xi^m \partial_m \Phi \,,
 }
where $\xi^m = (\xi^0, \xi^1, \xi^2, \xi^3, \xi^5)$ is the vector field parameterizing the gauge transformation.  In our coordinates $r = \Phi$, so $\widetilde{\delta \phi} = \delta \phi + \xi^5$.  Let's assume that initially $\delta \phi = 0$ and
 \eqn{OldPert}{
  \delta g_{mn} = \diag \left\{-h e^{2 A} H_{00}, e^{2A} H_{11}, e^{2A} H_{11}, e^{2A} H_{11}, e^{2B} H_{55}/h \right\} \,.
 }
We would like to perform a gauge transformation to axial gauge, where
 \eqn{NewPert}{
  \widetilde{\delta g}_{mn} = \diag \left\{-h e^{2 A} \widetilde{H}_{00}, e^{2A} \widetilde{H}_{11}, e^{2A} \widetilde{H}_{11}, e^{2A} \widetilde{H}_{11}, 0 \right\} \,,
 }
and where $\widetilde{\delta \phi} = \xi^5$ does not necessarily vanish.  The symmetries of the problem imply that $\xi^{1} = \xi^{2} = \xi^{3} = 0$, as well as $\xi^0 = \xi^0(t, \Phi)$ and $\xi^5 = \xi^5(t, \Phi)$.  To achieve the above gauge transformation, we need to solve the following two equations, coming from the $55$ and $50$ equations in \eqref{Gauge}, respectively:
 \eqn{xiEQs}{
  \partial_{\Phi} \xi^5 + B' \xi^5 - {h'\over 2h} \xi^5 + {H_{55} \over 2} &=0\cr
  \partial_{\Phi} \xi^0 - {e^{2 B - 2 A} \over h^2} \partial_t \xi^5 &= 0 \,.
 }
It follows from \eqref{Asymp} that near the boundary (i.e.~at small $\Phi$), $B' = -1/\Phi$ becomes much larger than $h'/2 h$, so the first equation in \eqref{xiEQs} has the asymptotic solution
 \eqn{xirSoln}{
  \widetilde{\delta \phi} \equiv \xi^5 \approx C(t) \Phi + \textrm{non-homogeneous solution} \,,
 }
where $C(t)$ is an integration constant that may depend on $t$.  By adjusting this constant, one can therefore construct a gauge transformation between \eqref{OldPert} and \eqref{NewPert} for which\footnote{We employ here the ``little $o$'' Landau symbol: $f(x) = o(g(x))$ as $x \to 0$ iff $\lim_{x \to 0} f(x)/g(x) = 0$.}
 $\widetilde{\delta \phi} = o(\Phi)$ as $\Phi\to 0$.  In addition, the $00$ and $11$ equations in \eqref{Gauge} give
 \eqn{Gottilde}{
  \widetilde{H}_{00} &= H_{00} + 2 A' \widetilde{\delta \phi} + {h' \over h} \widetilde{\delta \phi} + 2 \partial_t \xi^0 \cr
  \widetilde{H}_{11} &= H_{11} + 2 A' \widetilde{\delta \phi} \,,
 }
which, together with \eqref{xiEQs} and \eqref{Asymp}, also imply that
 $\widetilde{H}_{00} = H_{00} + o(\Phi)$ and $\widetilde{H}_{11} = H_{11} + o(\Phi)$ as $\Phi\to 0$, provided that $\Delta > 2$.  From $\widetilde{\delta \phi} = o(\Phi)$, together with $\widetilde{H}_{00} = H_{00} + o(\Phi)$ and $\widetilde{H}_{11} = H_{11} + o(\Phi)$, it follows that the operator dual to $H_{11}$ is the same in a gauge where $\delta \phi = 0$ and in axial gauge:  it is given by ${1\over 2} T_i^{\phantom{i}i}$.

Note that we might equally have chosen a gauge where $\widetilde{\delta \phi} \approx \alpha \Phi$ as $\Phi \to 0$ with non-zero $\alpha$.  In this gauge, $\tilde{H}_{00} = H_{00} + {2\over \Delta-4} \alpha + o(\Phi)$ and $\tilde{H}_{11} = H_{11} + {2\over \Delta-4} \alpha + o(\Phi)$.  The contribution to the deformation of the lagrangian proportional to $\alpha$ is then
 \eqn{alphaContrib}{
  \delta {\cal L} \propto {2\over \Delta - 4} \times {1 \over 2} T_\mu^{\phantom{\mu}\mu} + {\cal O}_\phi \,.
 }
This expression vanishes upon using the fact that $T_\mu^{\phantom{\mu}\mu} = -\beta {\cal O}_\phi = (4 - \Delta) {\cal O}_\phi$.

\section{Prescription for computing the retarded correlator}
\label{AppGR}

The prescription that we used for computing retarded correlators (equations \eqref{RetardedGreens} and \eqref{RetardedGreens2}) was proposed in \cite{Son:2002sd}, and it was justified in \cite{Herzog:2002pc} by using Schwinger-Keldysh contours.  In this section we will show how this prescription can also be justified through the fact that it provides an analytic continuation of the corresponding Euclidean correlator computed according to the prescription of \cite{Gubser:1998bc, Witten:1998qj}.  This fact was checked in \cite{Son:2002sd}, but a more general argument can be made.

For a bosonic operator ${\cal O}$, the Euclidean correlator is defined by
 \eqn{GEDef}{
  G_E(\omega_n, \vec{k}) = \int_0^\beta d\tau \int d^3 x\,
   e^{-i (\vec{k} \cdot \vec{x} - \omega_n \tau)}
   \langle T_E {\cal O}(\tau, \vec{x}) {\cal O}(0) \rangle \,,
 }
where $T_E$ denotes Euclidean time-ordering, $\tau$ is the Euclidean time, and $\omega_n = 2 \pi n / \beta$ are the Matsubara frequencies.  The Euclidean correlator is defined only at the Matsubara frequencies $\omega_n$ because in position space the time-ordered two-point function appearing in \eqref{GEDef} is periodic with period $\beta = 1/T$.  $G_E(\omega_n, \vec{k})$ is related to the retarded correlator $G_R(\omega, \vec{k})$ by analytic continuation:  $G_R(\omega, \vec{k})$, seen as a function of $\omega$, can be analytically continued to the upper half-plane, and for $\omega_n > 0$ one has
 \eqn{GRfromGE}{
  G_R(i \omega_n, \vec{k}) = - G_E(\omega_n, \vec{k}) \,.
 }

For simplicity, let's first examine the case where ${\cal O} = {\cal O}_\phi$ is dual to a massive bulk scalar field $\phi$ in the asymptotically-AdS background
 \eqn{phiBackground}{
  ds^2 = g_{\mu\nu} dx^\mu dx^\nu = e^{2A(r)} \left[ -h(r) dt^2 + d\vec{x}^2 \right] +
    e^{2B(r)} {dr^2 \over h(r)}
 }
with the conformal boundary at $r = 0$ and a black hole horizon with Hawking temperature $T$ at $r = r_H$.  The quadratic action for $\phi$ is
 \eqn{phiAction}{
  S = {1 \over 16 \pi G_5} \int d^4 x\, dr\, \sqrt{-g} \left[-{1\over 2} g^{\mu\nu} \partial_\mu \phi \partial_\nu \phi  -{1\over 2} m^2 \phi^2 \right] \,,
 }
and the equation of motion following from it is
 \eqn{phieom}{
  \square \phi - m^2 \phi = 0 \,.
 }
Defining the Fourier transform of $\phi$ by
 \eqn{phiFT}{
  \phi(\omega, \vec{k}, r)\equiv \int d^4 x \, e^{- i (\vec{k} \cdot \vec{x} - \omega t)} \phi(t, \vec{x}, r)\,,
 }
one can write \eqref{phieom} as
 \eqn{phieomFourier}{
  \left[\partial_r^2 + \left(4 A' - B' + {h'\over h} \right) \partial_r - {e^{2(B-A)} \over h} \left(-{\omega^2 \over h} + \vec{k}^2 + e^{2 A} m^2 \right) \right] \phi(\omega, \vec{k}, r) = 0 \,.
 }
For non-zero $\omega$, there are two linearly independent solutions to this equation:  one for which $\phi \approx |r_H-r|^{-i \omega / 4 \pi T}$ at $r \approx r_H$, corresponding to $\phi$-quanta falling into the black hole horizon, and one for which $\phi \approx |r_H-r|^{i \omega/ 4\pi T}$, corresponding to $\phi$-quanta coming out of the black hole horizon.  The expectation of \cite{Son:2002sd} was that the infalling solution should be the one related to the retarded correlator.  At the conformal boundary, a generic solution behaves as $\phi \approx e^{(\Delta - 4)A}$, where $\Delta$ is the more positive root of the equation
 \eqn{DeltaEQ}{
  \Delta (\Delta - 4) = m^2 L^2 \,.
 }

To go to Euclidean signature, one merely needs to replace $-h(r) dt^2$ by $h(r) d\tau^2$ in \eqref{phiBackground} and let $\tau$ run from $0$ to $\beta$.  The Euclidean metric is
 \eqn{phiBackgroundEuclidean}{
  ds_E^2 = g_{E, \mu\nu} dx^\mu dx^\nu = e^{2A(r)} \left[ h(r) d\tau^2 + d\vec{x}^2 \right] +
    e^{2B(r)} {dr^2 \over h(r)} \,,
 }
and the Euclidean action is
 \eqn{phiEuclideanAction}{
  S_E = {1 \over 16 \pi G_5} \int d^4 x\, dr\, \sqrt{g_E} \left[{1\over 2} g_E^{\mu\nu} \partial_\mu \phi_E \partial_\nu \phi_E  + {1\over 2} m^2 \phi_E^2 \right] \,.
 }
If we define the Fourier modes of $\phi_E$ by
 \eqn{phiEFourier}{
  \phi_E(\omega_n, \vec{k}, r) \equiv {1 \over \beta} \int_0^\beta d\tau \int d^3 x \, e^{-i (\vec{k} \cdot \vec{x} - \omega_n \tau)} \phi_E(\tau, \vec{x}, r) \,,
 }
then the equation of motion that follows from \eqref{phiEuclideanAction} is
 \eqn{phiEeomFourier}{
  \left[\partial_r^2 + \left(4 A' - B' + {h'\over h} \right) \partial_r - {e^{2(B-A)} \over h} \left({\omega_n^2 \over h} + \vec{k}^2 + e^{2 A} m^2 \right) \right] \phi_E(\omega_n, \vec{k}, r) = 0 \,.
 }
This equation is identical to \eqref{phieomFourier} except for the replacement $\omega^2 \to - \omega_n^2$.  For non-zero $\omega_n$, the near-horizon behavior of the two linearly independent solutions is $\phi_E \approx |r_H-r|^{|\omega_n| / 4 \pi T}$, corresponding to regular solutions, and $\phi_E \approx |r_H-r|^{- |\omega_n| / 4 \pi T}$, corresponding to divergent solutions.  The behavior of $\phi_E$ at the conformal boundary is the same as in the Minkowski case:  generically, $\phi_E \approx e^{(\Delta - 4)A}$, with $\Delta$ again defined as the more positive root of equation \eqref{DeltaEQ}.

The Euclidean correlator \eqref{GEDef} can be computed from
 \eqn{EuclideanADSCFT}{
  \langle e^{\int \phi_0 {\cal O}_\phi} \rangle = e^{-S_E^\textrm{on-shell}[\phi_E]} \,,
 }
where $\phi_0(\omega_n,\vec{k}) = \lim_{r\to 0} \phi_E(\omega_n,\vec{k},r) e^{(4 - \Delta) A}$ and it is assumed that $\phi_E$ is regular at the horizon \cite{Gubser:1998bc, Witten:1998qj}. The on-shell action $S_E^\textrm{on-shell}[\phi_E]$ can be obtained from integrating the first term in \eqref{phiEuclideanAction} by parts and using the equation of motion for $\phi_E$ to set the bulk term to zero and keep only the boundary term.  In Fourier space, the on-shell action can be expressed as
 \eqn{SEonshell}{
  S_E^{\rm on-shell} &= {\beta \over 32 \pi G_5} \sum_n \int {d^3 k \over (2 \pi)^3} J_E\left[\phi_E\right](\omega_n, \vec{k}, r) \Big|_{r = 0}^{r = r_H} \cr
    J_E\left[\phi_E\right](\omega_n, \vec{k}, r) &\equiv \phi_E(-\omega_n, -\vec{k}, r) h e^{4 A - B} \partial_r \phi_E(\omega_n, \vec{k}, r) \,.
 }
We now introduce the notation $\phi_E^{\rm reg}$ for a solution of \eqref{phiEeomFourier}
that is regular at the horizon and is normalized by $\phi_0=1$.
From \eqref{EuclideanADSCFT} it is straightforward to show that
 \eqn{GotGE}{
  G_E(\omega_n, \vec{k}) = -{1\over 32 \pi G_5} \left(
  J_E\left[\phi_E^{\rm reg}\right](\omega_n, \vec{k}, r) +
  J_E\left[\phi_E^{\rm reg}\right](-\omega_n, -\vec{k}, r) \right) \bigg|_{r = 0}^{r = r_H} \,.
 }
 Note that the equation of motion \eqref{phiEeomFourier} for $\phi_E$ as well as the boundary conditions are invariant under $\omega_n \to - \omega_n$, which implies that
 \eqn{JESymmetry}{
  J_E\left[\phi_E^{\rm reg}\right](\omega_n, \vec{k}, r) =
  J_E\left[\phi_E^{\rm reg}\right](-\omega_n, -\vec{k}, r) \,.
 }
So the four terms in \eqref{GotGE} are equal in pairs.  A further simplification occurs for $\omega_n \neq 0$:  starting from the fact that $\phi_E^{\rm reg} \approx |r_H-r|^{|\omega_n| / 4 \pi T}$ at $r \approx r_H$ and using the definition of $J_E$ in \eqref{SEonshell}, one can show that $J_E\left[\phi_E^{\rm reg}\right](\omega_n, \vec{k}, r) \approx |r_H-r|^{2 |\omega_n| / 4 \pi T} \to 0$ as $r \to r_H$.  So there is no contribution from the horizon in \eqref{GotGE}.  Consequently, for $\omega_n \neq 0$,
 \eqn{GotGESimp}{
  G_E(\omega_n, \vec{k}) = {1\over 16 \pi G_5} \lim_{r \to 0} J_E\left[\phi_E^{\rm reg}\right](\omega_n, \vec{k}, r) \,.
 }

In analogy to \eqref{SEonshell}, one can define
 \eqn{jDef}{
  S^{\rm on-shell} &= {1 \over 32 \pi G_5} \int {d\omega d^3 k \over (2 \pi)^4} J\left[\phi\right](\omega, \vec{k}, r)\Big|_{r = 0}^{r = r_H} \cr
   J\left[\phi\right](\omega, \vec{k}, r) &\equiv - \phi(-\omega, -\vec{k}, r) h e^{4 A - B} \partial_r \phi(\omega, \vec{k}, r) \,,
 }
which can be obtained from \eqref{phiAction} after integrating by parts and using the equation of motion \eqref{phieom} to set the bulk term to zero.  We now introduce the notation
$\phi^{\rm in}$ for a solution of \eqref{phieomFourier} that is infalling and is normalized by setting $\lim_{r\to0} \phi^{\rm in}(\omega,\vec{k},r) e^{ (4 -\Delta)A} = 1$ and
$\phi^{\rm out}$ for a solution that is outgoing and has the same boundary normalization.
From comparing equations \eqref{phieomFourier} and \eqref{phiEeomFourier} one can
easily see that if $\omega_n>0$ then $\phi_E^{\rm reg}(\omega_n)=\phi^{\rm in} (i\omega_n)$,
while if $\omega_n<0$ then $\phi_E^{\rm reg}(\omega_n)=\phi^{\rm out} (i\omega_n)$.
Because the behavior near the boundary of $\phi_E^{\rm reg}$, $\phi^{\rm in}$, and $\phi^{\rm out}$
is, by definition, the same regardless of the value of $\omega$ or $\omega_n$, it follows that
\eqn{JEandJ}{
  \lim_{r\to 0} J_E\left[\phi_E^{\rm reg}\right](\omega_n, \vec{k}, r)  =  \left\{
  \begin{array}{ll}
    - \lim_{r\to 0} J\left[\phi^{\rm in} \right](i \omega_n, \vec{k}, r) & \quad {\rm if}\quad\omega_n>0 \\
    - \lim_{r\to 0} J\left[\phi^{\rm out} \right](i \omega_n, \vec{k}, r) & \quad {\rm if}\quad\omega_n<0 \,.
  \end{array} \right.
}
Because \eqref{JEandJ} and \eqref{GotGESimp} imply \eqref{GRfromGE}, it follows by analytic continuation\footnote{The analytic continuation is unique if one further assumes that $G_R(\omega, \vec{k})$ decays fast enough at large complex $\omega$ \cite{Landauvol9}.} that $G_R$ should be given by
 \eqn{GRPrescription}{
  G_R(\omega, \vec{k}) = {1\over 16 \pi G_5} \lim_{r \to 0} J\left[\phi^{\rm in} \right](\omega, \vec{k}, r) \,.
 }
This prescription is exactly the one proposed in \cite{Son:2002sd}.\footnote{Note that if we wish to compute the advanced correlator we need only replace $\phi^{\rm in}$ by $\phi^{\rm out}$. If we denote the result of following this alternate prescription by $G_A(\omega,\vec{k})$, then it follows from \eqref{JEandJ} that $G_A(i \omega_n,\vec{k}) = - G_E(\omega_n,\vec{k})$ for $\omega_n < 0$, which is the analog of \eqref{GRfromGE} obeyed by the advanced correlator.}

The argument leading to \eqref{GRPrescription} was based on the assumption of non-vanishing $\omega$ and $\omega_n$.  For $\omega = \omega_n = 0$, the equation of motion for $\phi$ coincides with the one for $\phi_E$.  One of its linearly independent solutions approaches a constant at the horizon, and the other one diverges logarithmically as a function of $r_H-r$.  In this case, the Euclidean correlator may pick up a contribution from the horizon, which would mean an added contribution to \eno{GotGESimp}.  But by continuity, we expect \eno{GRPrescription} to apply unaltered at $\omega=0$  if $\phi^{\rm in}$ is replaced by the solution which is regular at the horizon.

The argument given above can be generalized to operators with higher spin, or to operators that mix with one another.  In the case of the shear viscosity computation in section~\ref{SOLVESHEAR}, one can set ${\cal O} = T_{12}$ and $\phi = H_{12}$.  In the case of the bulk viscosity computation, there are some additional subtleties. To avoid complications present in the general case, in the remainder of this appendix we will consider only perturbations that do not depend on the $\vec{x}$ coordinates, which is to say we set $\vec{k}=0$. The fields $H_{11}=H_{22}=H_{33}$, whose dual operator ${\cal O}={1 \over 2} T_i{}^i$ we are interested in, couple to other fields, and so we must replace $\phi(t,\vec{x},r)$ in the above discussion by $\vec{h}(t,r) = \begin{pmatrix} H_{00}(t,r) & H_{11}(t,r) & H_{55}(t,r) \end{pmatrix}$, as in section~\ref{BULKPERTURB}.  Let's work out how the Green's function prescription \eno{GRPrescription} generalizes to this case.

To compute two-point correlators, we need only consider the quadratic part of the action for $\vec{h}$. In terms of the Fourier transform
\eqn{hFourier}{
\vec{h}(\omega,r) = \int_{-\infty}^\infty {d\omega\over 2\pi}
  e^{i\omega t} \vec{h}(t,r) \,,
}
the quadratic action is
\eqn{phiGenAction}{
S = {1\over 32 \pi G_5} \int {d\omega \over 2 \pi}
   dr  \left(
  \partial_r\vec{h}^{*T} {\bf m} \partial_r\vec{h} +
  \vec{h}^{*T} {\bf b}^{*T}  \partial_r\vec{h} +
  \partial_r \vec{h}^{*T} {\bf b} \vec{h} +
  \vec{h}^{*T} {\bf k} \vec{h} \right) \,,
}
where $\bf{m}$, $\bf{b}$ and $\bf{k}$ are the matrix-valued functions of $\omega$ and $r$ given in \eqref{mkbDef}.  In order to obtain an explicit final result, one would need to evaluate the contribution of the symmetric matrix ${\bf G}$ described below \eno{MPhiDef}.  We will not do this here, but we believe it is straightforward in principle.

Integrating by parts and using the equations of motion that follow from \eqref{phiGenAction}, one can see that
\eqref{jDef} generalizes to
\eqn{GenSonshell}{
 S^{\rm on-shell} &= {1 \over 16 \pi G_5}
   \int_{-\infty}^\infty {d\omega \over 2 \pi}
    J\left[\vec{h}\right](\omega, r)\Big|_{r = 0}^{r = r_H} \cr
    J\left[\vec{h}\right](\omega, r) &\equiv \vec{h}^{T}(-\omega,  r)
    \left( {\bf m} \partial_r + {\bf b}  \right)
    \vec{h}(\omega, r) \,.
}
The Euclidean action $S_E$ is obtained from the real space counterpart of $\eqref{phiGenAction}$
by performing the Wick rotation $t \to -i\tau$ and adding an overall minus sign.  In Fourier space, the Wick rotation can be implemented by sending
$\omega$ to $i\omega_n$ and replacing the integral over $\omega$ by a discrete sum over $n$.  One can thus write
\eqn{JEGeneral}{
S_E^{\rm on-shell} &= {\beta \over 32 \pi G_5} \sum_n J_E\left[\vec{h}_E\right](\omega_n, r)\Big|_{r = 0}^{r = r_H} \cr
J_E\left[\vec{h}_E\right](\omega_n,  r) &\equiv \vec{h}_E^{T}(-\omega_n,  r)
   \left( {\bf m}_E \partial_r + {\bf b}_E \right)
   \vec{h}_E(\omega_n, r)
}
where ${\bf m}_E(\omega_n,r)=-{\bf m}(i\omega_n,r)$, and ${\bf b}_E$ and ${\bf k}_E$ obey similar relations.

The solutions of the equations of motion that follow from \eqref{GenSonshell} behave like infalling or outgoing waves near the horizon, namely $\vec{h} \approx |r_H - r|^{\pm i \omega/ 4\pi T}$ for $r \approx r_H$. Here, in a slight abuse of notation, this asymptotic form is understood to hold for every component of $\vec{h}$.  It follows from this that at $r\approx r_H$, the equations of motion for $h_E$ have divergent solutions that behave as $|r_H-r|^{- |\omega_n|/4\pi T}$ and regular solutions that behave as $|r_H-r|^{ |\omega_n|/4\pi T}$.

To compute correlators of the operator dual to $H_{11}$,  we must demand that $H_{00} \to 0$, $H_{11} \to 1$ and $H_{55}\to 0$ as $r \to 0$. This normalization is not affected by Wick rotation and will be the same for the Euclidean fields $\vec{h}_E$. It is useful to introduce the notation $\vec{h}_E^{\rm reg}$ for a solution of the equations of motion for $\vec{h}_E$ that is regular at the horizon and has this  normalization near the boundary. We also will use $\vec{h}^{\rm in}$ and $\vec{h}^{\rm out}$ to denote solutions of the equations of motion for $\vec{h}$ that are infalling or outgoing, respectively, and that are correctly normalized near the boundary.

Equation \eqref{GotGE} can be generalized to
\eqn{GotGEGen}{
  G_E(\omega_n, \vec{k}=0) = -{1\over 32 \pi G_5} \left(
  J_E\left[\vec{h}_E^{\rm reg}\right](\omega_n, r) +
  J_E\left[\vec{h}_E^{\rm reg}\right](-\omega_n, r) \right) \bigg|_{r = 0}^{r = r_H} \,.
 }
It is easy to see that $J_E\left[\vec{h}_E\right](-\omega_n,r)=J_E\left[\vec{h}_E\right](\omega_n,r)$ and that $J_E\left[\vec{h}_E^{\rm reg}\right](\omega_n,r_H)=0$, and so
\eqn{GotGEGenSimp}{
  G_E(\omega_n, \vec{k}=0) = -{1\over 16 \pi G_5} \lim_{r\to 0}  J_E\left[\vec{h}_E^{\rm reg}\right](\omega_n, r) \,.
 }
Finally, it is clear that $\vec{h}^{\rm in}(i \omega_n,r)=\vec{h}_E^{\rm reg}(\omega_n,r)$ for $\omega_n>0$ and $\vec{h}^{\rm out}(i \omega_n,r)=\vec{h}_E^{\rm reg}(\omega_n)$ for $\omega_n<0$.  The same reasoning as in the case of the bulk scalar field then shows that \eqref{GRPrescription} can be generalized to
 \eqn{GRPrescriptionGen}{
  G_R(\omega,\vec{k}=0) = {1\over 16 \pi G_5} \lim_{r \to 0} J\left[\vec{h}^{\rm in} \right](\omega, r) \,.
 }

The imaginary part of the retarded correlator can be expressed in terms of a conserved flux ${\cal F}$.  Indeed, \eqref{phiGenAction} is invariant under $\vec{h} \to e^{i \alpha} \vec{h}$ and the conserved quantity associated to this symmetry is exactly
\eqn{GenFluxDef}{
  {\cal F}\left[\vec{h}\right](\omega) = - \Im J\left[\vec{h}\right](\omega, r) \,.
}
Since ${\cal F}$ is independent of $r$, it follows from \eqref{GRPrescription} that $\Im G_R = - {{\cal F} \over 16 \pi G_5}$.

\bibliographystyle{ssg}
\bibliography{sound}
\end{document}